\documentclass{article}

\PassOptionsToPackage{numbers, compress}{natbib}

\usepackage[final]{neurips_2020}

\usepackage[utf8]{inputenc} 
\usepackage[T1]{fontenc}    
\usepackage[hyphens]{url}            
\usepackage{hyperref}       
\usepackage{booktabs}       
\usepackage{amsfonts}       
\usepackage{nicefrac}       
\usepackage{microtype}      

\title{Non-portability of Algorithmic Fairness in India}

 \author{
   Nithya Sambasivan, Erin Arnesen, Ben Hutchinson, Vinodkumar Prabhakaran \\
  Google Research\\
  USA\\
  \texttt{(nithyasamba, erinarnesen, benhutch, vinodkpg)@google.com} \\
 }

\usepackage{comment}
\usepackage{graphicx}
\usepackage{tabularx}
\usepackage{enumitem}
\usepackage{color_edits}
\usepackage{wrapfig}

\begin{document}
\newcommand{\etal}{\textit{et al. \@}}
\newcommand{\eg}{\textit{e.g., \@}}
\newcommand{\ie}{\textit{i.e., \@}}
\newcommand{\toadd}[1]{{\color{red}VP: #1}}
\newcommand{\vinod}[1]{{\color{red}VP: #1}}
\newcommand{\ben}[1]{{\color{blue}BH: #1}}
\addauthor{BH}{blue}
\newcommand{\nithya}[1]{{\color{red}NS: #1}}

\newcommand{\vpedit}[1]{{\color{red}#1}}

\newcommand{\myfigureshrinker}{\vspace{-1cm}}

\newcommand{\tabitem}{\hspace{2pt}\textbullet~~}

\maketitle

\begin{abstract}
Conventional algorithmic fairness is Western in its sub-groups, values, and optimizations. In this paper, we ask how portable the assumptions of this largely Western take on algorithmic fairness are to a different geo-cultural context such as India. Based on 36 expert interviews with Indian scholars, and an analysis of emerging algorithmic deployments in India, we identify three clusters of challenges that engulf the large distance between machine learning models and oppressed communities in India. We argue that a mere translation of technical fairness work to Indian subgroups may serve only as a window dressing, and instead, call for a collective re-imagining of Fair-ML, by re-contextualising data and models, empowering oppressed communities, and more importantly, enabling ecosystems. 
\end{abstract}

\vspace{-.7em}
\section{Introduction}
\label{sec_intro}

Fairness research in Machine Learning (ML) has seen rapid growth in recent years; however, it remains largely rooted in Western concerns and histories: the injustices they focus on (\eg{}along race and gender), the datasets they study (\eg{}ImageNet), the measurement scales they use (\eg{}Fitzpatrick), and the legal tenets they draw from (\eg{}equal opportunity). This implicitly Western take is troublingly becoming a \textit{universal} ethical framework for ML, \eg{}AI strategies from India \cite{NITI}, Tunisia \cite{TunisiaAI}, and Mexico \cite{MexicoAI} all derive from this work, but fails to account for the several assumptions conventional algorithmic fairness makes about the availability and efficacy of the surrounding institutions and infrastructures. However, these infrastructures, values, and legal systems cannot be naively generalised to various non-Western countries. 

Let us consider the example of facial recognition technology, where demonstration of fairness failures resulted in bans and moratoria in the US. Several factors led to this outcome: 
decades of empiricism on proxies and metrics that correspond to subgroups in the West \cite{fitzpatrick1988validity}; public datasets, APIs, and laws enabling analysis of model outcomes \cite{MachineB33:online, HowFacia39:online}; an ML research/industry responsive to bias reports from users and civil society \cite{Weareimp72:online, buolamwini2018gender}; the existence of government representatives glued into technology policy \cite{JayapalJ44:online}; and an active media that scrutinizes downstream impacts of AI \cite{HowFacia39:online}. However, due to various cultural, ethnic, and infrastructural differences, these factors are often absent or irrelevant in much else of the world. Could fairness have structurally different meanings or mechanisms in non-Western contexts? How do social, economic, and infrastructural factors influence implementation of meaningful fairness?

In this position paper, we present insights from qualitative interviews with 36 Indian scholars and activists, and a discourse analysis of emerging algorithmic deployments in India. India is home to 1.38 billion people and their multiple languages, religions, cultural systems, and ethnicities. AI deployments are prolific in the public sector, \eg{}in predictive policing \cite{PredictiveIndia}, facial recognition \cite{FacialRecIndia}, and agriculture \cite{AgricIndia}. Despite this forward momentum, there is a dire lack of conversations on advancing algorithmic fairness for such a large population. Our paper presents a first step towards filling this important gap. We contend that conventional Fair-ML approach may be inappropriate and inimical in India, if it does not engage with local structures. For a country as deeply plural, complex, and contradictory as India---where the distance between models and oppressed communities is large---optimising model fairness alone can be mere tokenism. We call for action along three critical pathways towards algorithmic fairness in India: \textit{Recontextualising}, \textit{Empowering}, and \textit{Enabling} (see Figure~\ref{fig:Indiaframework}). The considerations we present involve collective responsibility of inter-disciplinary Fair-ML researchers, and push the bounds of what is considered to be fairness research. 

\vspace{-.7em}
\section{Methodology}
\label{sec_methodology}

We perform a synthesis of (i) 36 interviews with researchers and activists across various disciplines working with marginalized Indian communities, and (ii) observations of current algorithmic deployments and policies in India. We chose our respondents from diverse disciplines: Computer Science (11), Activism (9), Law and Public Policy (6), Science and Technology Studies (5), Development Economics (2), Sociology (2), and Journalism (1). All respondents had 10-30 years of experience working with on social justice in India, on areas including caste, gender, labour, disability, surveillance, health, and constitutional rights. In conjunction with qualitative interviews, we analyzed various algorithmic deployments and emerging policies in India, starting from 2009. We identified and analysed various Indian news publications, policy documents, grassroots fora, and prior research on automation in India. The systematic process followed for the collection, categorization, and synthesis of data to develop the key insights below are described in detail in Appendix~\ref{app_methodology}. 

\vspace{-.7em}
\section{Summary of Findings}
\label{sec_findings}

\begin{wrapfigure}{r}{0.47\textwidth}
\vspace{-.7cm}
  \begin{center}
    \includegraphics[width=\linewidth]{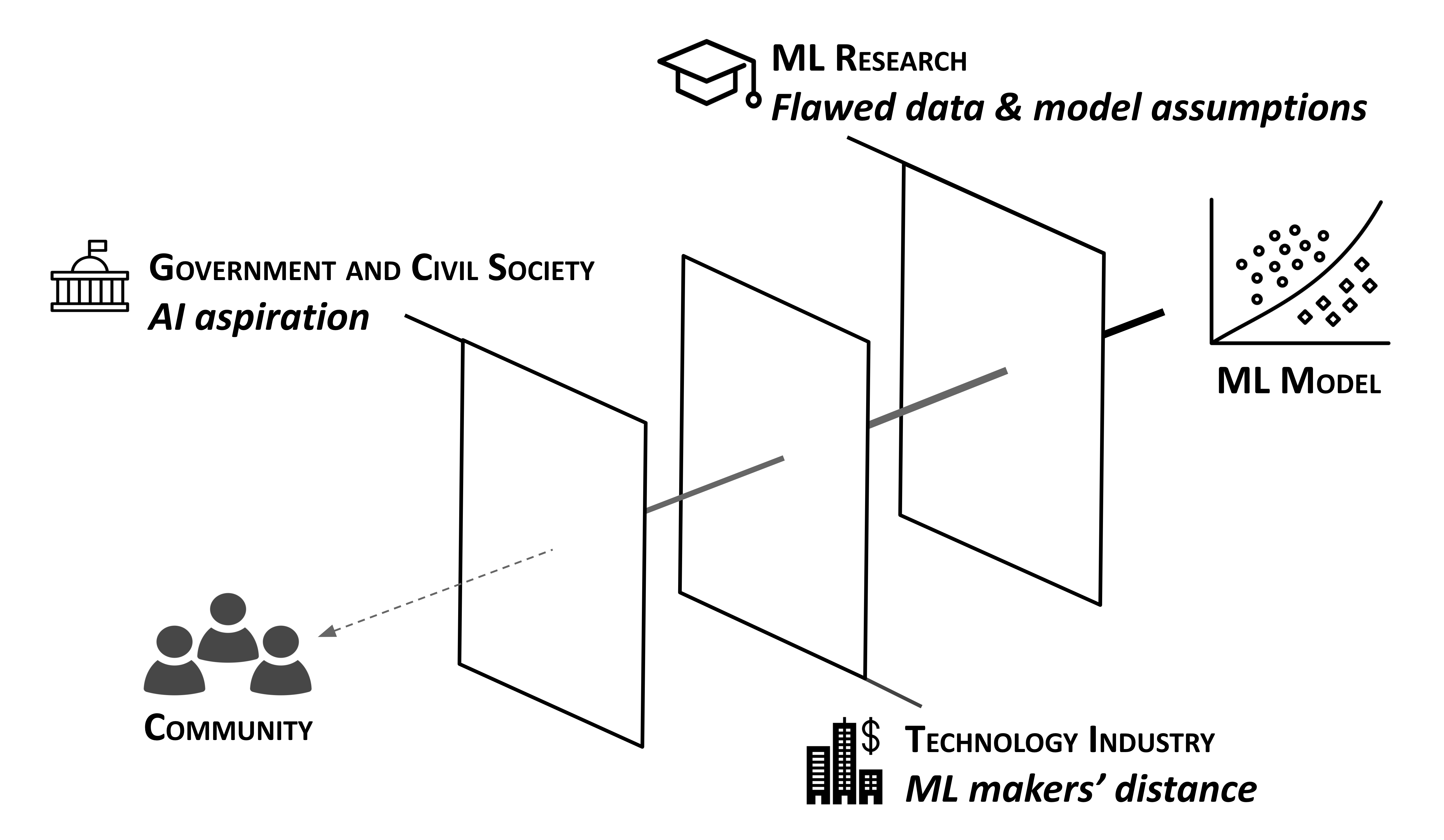}
  \end{center}
\caption{\label{fig:Indiadistance} \small Challenges to algorithmic fairness in India: A Model-Community distance perspective}
\vspace{-.8cm}
\end{wrapfigure}

Our study finds that the conventional fairness approach may be inappropriate, insufficient, or even inimical in India, if it does not engage with local structures. In India where the distance between ML models and dis-empowered communities whom they aim to serve is large---via technical distance, social distance, ethical distance, temporal distance, and physical distance---a myopic take on localising `fair' model outputs alone can backfire. We identify three high-level themes/clusters of challenges (Figure~\ref{fig:Indiadistance}) that engulf the large distance between models deployed in India and the communities they impact. 
We summarize them below.

\textbf{Flawed data and model assumptions}: We found numerous ways data and model assumptions fail in India, primarily: \textit{data distortions} owing to infrastructural challenges and technology usage patterns (\eg{}SIM card sharing among women \cite{sambasivan2018privacy}) that makes datasets not faithful representations of people and phenomena; \textit{data incompleteness} due to the models gearing for the `good' data profiles that largely favor middle-class men---in a society where 50\% do not have Internet, especially women and lower caste people; \textit{axes of discrimination} and their proxies and indicators that differ from how they manifest in the West (for a detailed account, see Appendix~\ref{app_subgroups_table}); and \textit{lack of Indic justice approaches} like reservations/quotas that present new challenges to fairness evaluations and interventions. 

\textbf{ML makers' distance}: Several respondents described a transactional mindset towards Indians, seeing them
as agency-less data subjects that generated large-scale behavioural traces to improve ML models (also see \cite{Digitalc91:online}), resulting in poor recourse and redress for Indian users, especially for those dis-empowered by caste, religion, or gender. The human infrastructures who played a crucial role in providing recourse to marginalised Indian communities (\eg{}street-side bureaucrats, and front line workers) were removed in ML systems. Many respondents also pointed out the lack of diverse representation in Indian tech workforce that the distant ML makers often fail to recognize. 

\textbf{AI aspiration}: AI is aspirationally viewed in India, with high reliance for high-stakes domains, accompanied with high trust in automation, limited transparency and the lack of an empowered ecosystem to question algorithmic interventions. For instance, respondents pointed to the lack of a buffer zone of journalists, activists, and researchers to keep ML system builders accountable. Lack of transparency and stakeholder participation, along with the its ‘neutral’ and ‘human-free’ associations lent misplaced credence to its algorithmic authority, making it further inscrutable.

\vspace{-.7em}
\section{Towards Algorithmic Fairness in India}

\begin{figure*}[t!]
\centering
\includegraphics[width=.86\linewidth]{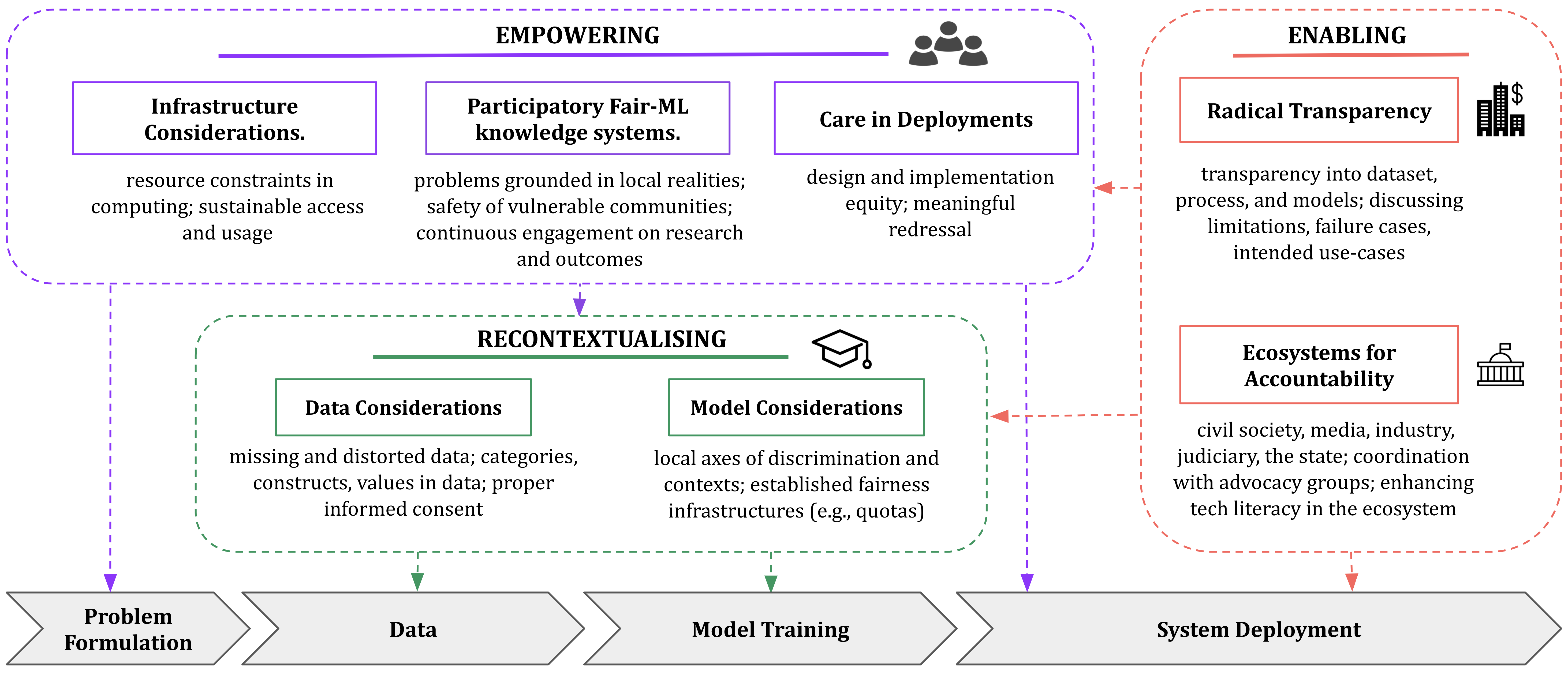}
\caption{Research pathways for Algorithmic Fairness in India.}
\label{fig:Indiaframework}
\end{figure*}

To account for the challenges outlined above, we need to understand and design for \textit{end-to-end} chains of algorithmic power, including how AI systems are conceived, produced, utilised, appropriated, assessed, and contested in India. To this end, we propose a research agenda where we call for action along three critical and contingent pathways towards successful AI fairness in India: \textit{Recontextualising}, \textit{Empowering}, and \textit{Enabling} (see Figure~\ref{fig:Indiaframework}). These pathways present new sociotechnical research challenges and require cross-disciplinary and cross-institutional collaborations. 

\vspace{-.7em}
\subsection{Recontextualising data and models}

\textbf{Data} plays a critical role in measurements and mitigation of algorithmic bias. We must be (even more than usual) sceptical of Indian datasets (due to challenges outlined above), and study how fairness research could handle the known distortions in the data as well as how we must account for data voids \cite{golebiewski2019data} for which statistical extrapolations might be invalid. The vibrant role played by human infrastructures in providing, negotiating, collecting, and stewarding data points should encourage us to re-imagine data as a dialogue rather than operations, i.e, products of both the beholder and the beheld. How might data consent work in this context? One approach could be to create transitive informed consent, built upon personal relationships and trust in data workers, with transparency on potential downstream applications. Ideas like collective consent \cite{MozillaF26:online}, data trusts \cite{Nesta}, and data co-ops may enhance community agency in datasets, while simultaneously increasing data reliability. Finally, we must question the categories and constructs we model in datasets, and how we measure them. When categories are appropriate for endemic goals (\eg{} caste membership for quotas), what form should their distributions and annotations take? Linguistically and culturally pluralistic communities should be given voices in these negotiations in ways that respect Indian norms of representation.

\textbf{Model} (un)fairness detection and mitigation should incorporate the prominent axes of historical injustices in India (see Appendix~\ref{app_subgroups_table}) and tackle the challenges in operationalising them for testing AI; \eg{}representational biases of caste and other sub-groups in NLP models, biases in Indic language NLP including challenges from code-mixing, Indian subgroup biases in computer vision, tackling online misinformation, benchmarking using Indic datasets, and fair allocation models in public welfare. It is important to note that operationalising fairness approaches from the West to these axes is often nontrivial. For instance, personal names act as a signifier for various socio-demographic attributes in India, however there are no large datasets of Indian names (like the US Census data, or the SSA data) that are readily available for fairness evaluations. Another important consideration is how the algorithmic fairness interventions work with the existing infrastructures in India that surrounds decision making processes; \eg{}in the context of restorative justice initiatives such as reservations/quotas? The quota system gives rise to the problem of matching under distributional constraints \cite{kamada2015efficient, goto2017designing, ashlagi2020assignment} that has not received much attention within the ML research. 

\vspace{-.7em}
\subsection{Empowering communities}

%
\textbf{Participatory Fair-ML knowledge systems} are critical to effectively fill the distance between ML makers and communities.
Context-free assumptions in fairness research, whether in homegrown or international ML systems, can not just fail, but produce harm inadvertently when applied to different infrastructural and cultural contexts. Systems should be created that enable marginalised communities to participate in the production of knowledge systems about themselves in ML frameworks. Grassroots efforts like Deep Learning Indaba \cite{dlindaba} and Khipu \cite{khipu} are helpful examples in bootstrapping AI research in communities. Initiatives like Design Beku \cite{designbeku} and SEWA \cite{sewa} are good decolonial examples of participatorily co-designing with under-served communities. Diverse values in algorithmic fairness can come from justice systems of indigenous Adivasis \cite{xaxa2011tribes}, \textit{dharma} (social ethic) and \textit{karma} (action) from Hinduism (note that Hindu concepts are sometimes rejected by Dalits for reinforcing hierarchy \cite{narayanan2001water}) or (\textit{sanghas}) (communism) espoused by Dr. B.R. Ambedkar \cite{ambedkar2011buddha}, for example.

\textbf{Low-resource considerations} pertaining to India should guide ML researchers to think beyond making model outputs fair, and about India's heterogeneous literacies, economics, and infrastructures that interfere with access to those fair outputs. Half the population of India is not online. Layers of the stack like interfaces, devices, connectivity, languages, and costs are all important in ensuring sustainable access and usage. Tackling resource constraints in computing disciplines like \textsc{ICTD} \cite{toyama2015geek} and \textsc{HCI4D} \cite{dray2012human} can help---where constraints have been embraced as design material, \eg{}delay-tolerant connectivity, low cost devices, text-free interfaces, partnership with civil society, and capacity building \cite{brewer2005case, heimerl2013local, kumar2015mobile, medhi2006text, sambasivan2019they}. Data infrastructures to build, share and sustain localised datasets would also enhance access equity to holding AI accountable (\eg{}\cite{lacuna}).

\textbf{Care in deployments} should be a primary concern.
Critiques were raised in our study on how neo-liberal ML followed a broader pattern of extraction from the `bottom billion' data subjects and AI labourers. Low costs, large and diverse populations, and policy infirmities have been cited as reasons for following double standards in India, \eg{}in non-consensual human trials and waste dumping \cite{macklin2004double} (also see \cite{mohamed2020decolonial}). Past disasters in India, like the fatal Union Carbide gas leak in 1984---one of the world's worst industrial accidents---point to faulty design and low quality standards for the `third world' \cite{Courttol47:online}. Similarly, overgrowth of ML capital, with unequal standards, inadequate safeguards, and dubious applications can lead to catastrophic effects (similar analogies have been made for content moderation \cite{roberts2016digital, sambasivan2019they}). ML researchers should study how to ensure meaningful recourse within the ecosystems they are embedding, and respond and engage with Indian realities and user feedback.

\vspace{-.7em}
\subsection{Enabling fair ML ecosystems}

\textbf{Ecosystems for accountability} requires enabling the civil society, media, industry, judiciary, and the state to meaningfully engage in holding AI interventions responsible. Moving from ivory tower research approaches to solidarity with various stakeholders through project partnerships, involvement in evidence-based policy, and policy maker education can help create a sustainable Fair-ML ecosystem based on sound empirical and ethical basis. Investigative journalism on algorithms is lacking in India, but sustainable partnerships may be created with rigorous media sources. Algorithmic fairness and ethics are not mainstream research topics in Indian academia today, but the academy has a pivotal role in advancing fairness in India. A concerted effort is required to create APIs, documentation, and socio-economic datasets to enable meaningful and equitable accountability in the ecosystem. 

\textbf{Radical transparency} is required to counteract the inscrutability and authority created by the enthusiastic AI aspiration. Besides the role played by ecosystem-wide regulations and standards, radical transparency as a design principle should be espoused and enacted by Fair-ML researchers committed to India. Transparency on datasets, processes, and models (\eg{}\cite{mitchell2019model,gebru2018datasheets,bender2018data}), openly discussing limitations, failure cases, and intended use-cases can help bolster the practical limits to applying computing to human problems. It is also incumbent upon those in power to discuss current Fair-ML interventions as a flawed and evolving approach. Such approaches can help move from the `magic pill' role of fairness as a checklist for ethical issues in India, to a more pragmatic function.

\vspace{-0.7em}
\section{Conclusion}

We are currently facing a watershed moment for AI, with extravagant hype and pervasive applications, with nations being turned into AI superpowers and rulers \cite{Whoeverl94:online}. We bring to light how efforts in the West to make AI fairer may not transfer readily elsewhere, specifically in the case of India, through an empirical bottom-up analysis. We present a roadmap to go beyond localising data and model fairness alone, and present a holistic research agenda to operationalize Fair-ML in India. While it is important to incorporate Indian concepts of fairness into ML that impacts Indian communities, the broader call is to develop general approaches to fairness that reflect local considerations.

\small
\bibliographystyle{plainnat}
\bibliography{references}

\begin{thebibliography}{54}
\providecommand{\natexlab}[1]{#1}
\providecommand{\url}[1]{\texttt{#1}}
\expandafter\ifx\csname urlstyle\endcsname\relax
  \providecommand{\doi}[1]{doi: #1}\else
  \providecommand{\doi}{doi: \begingroup \urlstyle{rm}\Url}\fi

\bibitem[NIT(2018)]{NITI}
\emph{National Strategy for Artificial Intelligence \#AI4ALL}.
\newblock Niti Aayog, 2018.

\bibitem[des(2020)]{designbeku}
\emph{Design Beku}, 2020.
\newblock URL \url{https://designbeku.in/}.

\bibitem[dli(2020)]{dlindaba}
\emph{Deep Learning Indaba}, 2020.
\newblock URL \url{https://deeplearningindaba.com/2020/}.

\bibitem[khi(2020)]{khipu}
\emph{Khipu AI}, 2020.
\newblock URL \url{https://github.com/khipu-ai}.

\bibitem[lac(2020)]{lacuna}
\emph{Lacuna Fund}, 2020.
\newblock URL \url{https://lacunafund.org/}.

\bibitem[sew(2020)]{sewa}
\emph{SEWA}, 2020.
\newblock URL \url{http://www.sewa.org/}.

\bibitem[Abraham and Rao()]{84DeadIn98:online}
Delna Abraham and Ojaswi Rao.
\newblock 84\% dead in cow-related violence since 2010 are muslim; 97\% attacks
  after 2014 | indiaspend.
\newblock
  \url{https://archive.indiaspend.com/cover-story/86-dead-in-cow-related-violence-since-2010-are-muslim-97-attacks-after-2014-2014}.
\newblock (Accessed on 08/16/2020).

\bibitem[Amazon(2020)]{Weareimp72:online}
Amazon.
\newblock We are implementing a one-year moratorium on police use of
  rekognition.
\newblock
  \url{https://blog.aboutamazon.com/policy/we-are-implementing-a-one-year-moratorium-on-police-use-of-rekognition},
  June 2020.
\newblock (Accessed on 08/29/2020).

\bibitem[Ambedkar(2011)]{ambedkar2011buddha}
Bhimrao~Ramji Ambedkar.
\newblock \emph{The Buddha and his dhamma: a critical edition}.
\newblock Oxford University Press, 2011.

\bibitem[Ambedkar(2014)]{ambedkar2014annihilation}
Bhimrao~Ramji Ambedkar.
\newblock \emph{Annihilation of caste: The annotated critical edition}.
\newblock Verso Books, 2014.

\bibitem[Angwin et~al.(2016)Angwin, Larson, Mattu, and
  Kirchner]{MachineB33:online}
Julia Angwin, Jeff Larson, Surya Mattu, and Lauren Kirchner.
\newblock Machine bias — propublica.
\newblock
  \url{https://www.propublica.org/article/machine-bias-risk-assessments-in-criminal-sentencing},
  May 2016.
\newblock (Accessed on 07/30/2020).

\bibitem[Ashlagi et~al.(2020)Ashlagi, Saberi, and
  Shameli]{ashlagi2020assignment}
Itai Ashlagi, Amin Saberi, and Ali Shameli.
\newblock Assignment mechanisms under distributional constraints.
\newblock \emph{Operations Research}, 68\penalty0 (2):\penalty0 467--479, 2020.

\bibitem[Baxi(2018)]{PredictiveIndia}
Abhishek Baxi.
\newblock Law enforcement agencies in india are using artificial intelligence
  to nab criminals.
\newblock
  \url{https://www.forbes.com/sites/baxiabhishek/2018/09/28/law-enforcement-agencies-in-india-are-using-artificial-intelligence-to-nab-criminals-heres-how},
  September 2018.
\newblock (Accessed on 08/30/2020).

\bibitem[Bender and Friedman(2018)]{bender2018data}
Emily~M Bender and Batya Friedman.
\newblock Data statements for natural language processing: Toward mitigating
  system bias and enabling better science.
\newblock \emph{Transactions of the Association for Computational Linguistics},
  6:\penalty0 587--604, 2018.

\bibitem[Bogner et~al.(2009)Bogner, Littig, and Menz]{bogner2009interviewing}
Alexander Bogner, Beate Littig, and Wolfgang Menz.
\newblock \emph{Interviewing experts}.
\newblock Springer, 2009.

\bibitem[Brewer et~al.(2005)Brewer, Demmer, Du, Ho, Kam, Nedevschi, Pal, Patra,
  Surana, and Fall]{brewer2005case}
Eric Brewer, Michael Demmer, Bowei Du, Melissa Ho, Matthew Kam, Sergiu
  Nedevschi, Joyojeet Pal, Rabin Patra, Sonesh Surana, and Kevin Fall.
\newblock The case for technology in developing regions.
\newblock \emph{Computer}, 38\penalty0 (6):\penalty0 25--38, 2005.

\bibitem[Buolamwini and Gebru(2018)]{buolamwini2018gender}
Joy Buolamwini and Timnit Gebru.
\newblock Gender shades: Intersectional accuracy disparities in commercial
  gender classification.
\newblock In \emph{Conference on fairness, accountability and transparency},
  pages 77--91, 2018.

\bibitem[Central Statistics~Office and
  Implementation(2018)]{IndiaMinistryStatistics2018}
Ministry of~Statistics Central Statistics~Office and Programme Implementation.
\newblock Women and men in india: A statistical compilation of gender related
  indicators in india.
\newblock Technical report, Government of India, 2018.

\bibitem[Dixit(2019)]{FacialRecIndia}
Pranav Dixit.
\newblock India is creating a national facial recognition system.
\newblock
  \url{https://www.buzzfeednews.com/article/pranavdixit/india-is-creating-a-national-facial-recognition-system-and},
  October 2019.
\newblock (Accessed on 08/30/2020).

\bibitem[Dray et~al.(2012)Dray, Light, Dearden, Evers, Densmore, Ramachandran,
  Kam, Marsden, Sambasivan, Smyth, et~al.]{dray2012human}
Susan Dray, Ann Light, A~Dearden, Vanessa Evers, Melissa Densmore,
  D~Ramachandran, M~Kam, G~Marsden, N~Sambasivan, T~Smyth, et~al.
\newblock Human--computer interaction for development: Changing human--computer
  interaction to change the world.
\newblock In \emph{The Human-Computer Interaction Handbook: Fundamentals,
  Evolving Technologies, and Emerging Applications, Third Edition}, pages
  1369--1394. CRC press, 2012.

\bibitem[Fitzpatrick(1988)]{fitzpatrick1988validity}
Thomas~B Fitzpatrick.
\newblock The validity and practicality of sun-reactive skin types i through
  vi.
\newblock \emph{Archives of dermatology}, 124\penalty0 (6):\penalty0 869--871,
  1988.

\bibitem[Gebru et~al.(2018)Gebru, Morgenstern, Vecchione, Vaughan, Wallach,
  Daum{\'e}~III, and Crawford]{gebru2018datasheets}
Timnit Gebru, Jamie Morgenstern, Briana Vecchione, Jennifer~Wortman Vaughan,
  Hanna Wallach, Hal Daum{\'e}~III, and Kate Crawford.
\newblock Datasheets for datasets.
\newblock \emph{arXiv preprint arXiv:1803.09010}, 2018.

\bibitem[Golebiewski and Boyd(2019)]{golebiewski2019data}
Michael Golebiewski and Danah Boyd.
\newblock Data voids: Where missing data can easily be exploited.
\newblock \emph{Data \& Society}, 2019.

\bibitem[Goto et~al.(2017)Goto, Kojima, Kurata, Tamura, and
  Yokoo]{goto2017designing}
Masahiro Goto, Fuhito Kojima, Ryoji Kurata, Akihisa Tamura, and Makoto Yokoo.
\newblock Designing matching mechanisms under general distributional
  constraints.
\newblock \emph{American Economic Journal: Microeconomics}, 9\penalty0
  (2):\penalty0 226--62, 2017.

\bibitem[Heimerl et~al.(2013)Heimerl, Hasan, Ali, Brewer, and
  Parikh]{heimerl2013local}
Kurtis Heimerl, Shaddi Hasan, Kashif Ali, Eric Brewer, and Tapan Parikh.
\newblock Local, sustainable, small-scale cellular networks.
\newblock In \emph{Proceedings of the Sixth International Conference on
  Information and Communication Technologies and Development: Full
  Papers-Volume 1}, pages 2--12, 2013.

\bibitem[IDSN(2010)]{Twothird31:online}
IDSN.
\newblock Two thirds of india’s dalits are poor - international dalit
  solidarity network.
\newblock \url{https://idsn.org/two-thirds-of-indias-dalits-are-poor/}, July
  2010.
\newblock (Accessed on 08/13/2020).

\bibitem[Kamada and Kojima(2015)]{kamada2015efficient}
Yuichiro Kamada and Fuhito Kojima.
\newblock Efficient matching under distributional constraints: Theory and
  applications.
\newblock \emph{American Economic Review}, 105\penalty0 (1):\penalty0 67--99,
  2015.

\bibitem[Kofman(2016)]{HowFacia39:online}
Ava Kofman.
\newblock How facial recognition can ruin your life -- intercept.
\newblock
  \url{https://theintercept.com/2016/10/13/how-a-facial-recognition-mismatch-can-ruin-your-life/},
  October 2016.
\newblock (Accessed on 07/30/2020).

\bibitem[Kumar and Anderson(2015)]{kumar2015mobile}
Neha Kumar and Richard~J Anderson.
\newblock Mobile phones for maternal health in rural india.
\newblock In \emph{Proceedings of the 33rd Annual ACM Conference on Human
  Factors in Computing Systems}, pages 427--436, 2015.

\bibitem[Kwet()]{Digitalc91:online}
Michael Kwet.
\newblock Digital colonialism is threatening the global south | science \&
  technology | al jazeera.
\newblock
  \url{https://www.aljazeera.com/indepth/opinion/digital-colonialism-threatening-global-south-190129140828809.html}.
\newblock (Accessed on 08/29/2020).

\bibitem[Macklin(2004)]{macklin2004double}
Ruth Macklin.
\newblock \emph{Double standards in medical research in developing countries},
  volume~2.
\newblock Cambridge University Press, 2004.

\bibitem[Martinho-Truswell et~al.(2018)Martinho-Truswell, Miller, Asare,
  Petheram, Stirling, G\'{o}mez~Mont, and Martinez]{MexicoAI}
Emma. Martinho-Truswell, Hannah. Miller, Isak~Nti Asare, Andre Petheram,
  Richard (Oxford~Insights) Stirling, Constanza G\'{o}mez~Mont, and Cristina
  (C~Minds) Martinez.
\newblock Towards an ai strategy in mexico: Harnessing the ai revolution.
\newblock In \emph{AI whitepaper.}, 2018.

\bibitem[Medhi et~al.(2006)Medhi, Sagar, and Toyama]{medhi2006text}
Indrani Medhi, Aman Sagar, and Kentaro Toyama.
\newblock Text-free user interfaces for illiterate and semi-literate users.
\newblock In \emph{2006 international conference on information and
  communication technologies and development}, pages 72--82. IEEE, 2006.

\bibitem[Microsoft(2017)]{AgricIndia}
Microsoft.
\newblock Digital agriculture: Farmers in india are using ai to increase crop
  yields.
\newblock
  \url{https://news.microsoft.com/en-in/features/ai-agriculture-icrisat-upl-india/},
  November 2017.
\newblock (Accessed on 08/30/2020).

\bibitem[Ministry~of Home~Affairs()]{CensusIndia2011}
Government of~India Ministry~of Home~Affairs.
\newblock 2011 census data.
\newblock \url{https://www.censusindia.gov.in/2011-Common/CensusData2011.html}.
\newblock (Accessed on 08/26/2020).

\bibitem[Mitchell et~al.(2019)Mitchell, Wu, Zaldivar, Barnes, Vasserman,
  Hutchinson, Spitzer, Raji, and Gebru]{mitchell2019model}
Margaret Mitchell, Simone Wu, Andrew Zaldivar, Parker Barnes, Lucy Vasserman,
  Ben Hutchinson, Elena Spitzer, Inioluwa~Deborah Raji, and Timnit Gebru.
\newblock Model cards for model reporting.
\newblock In \emph{Proceedings of the conference on fairness, accountability,
  and transparency}, pages 220--229, 2019.

\bibitem[Mohamed et~al.(2020)Mohamed, Png, and Isaac]{mohamed2020decolonial}
Shakir Mohamed, Marie-Therese Png, and William Isaac.
\newblock Decolonial ai: Decolonial theory as sociotechnical foresight in
  artificial intelligence.
\newblock \emph{Philosophy \& Technology}, pages 1--26, 2020.

\bibitem[Mukherji()]{TheCisco63:online}
Anahita Mukherji.
\newblock The cisco case could expose rampant prejudice against dalits in
  silicon valley.
\newblock
  \url{https://thewire.in/caste/cisco-caste-discrimination-silicon-valley-dalit-prejudice}.
\newblock (Accessed on 08/14/2020).

\bibitem[Mulgan and Straub()]{Nesta}
Geoff Mulgan and Vincent Straub.
\newblock The new ecosystem of trust: how data trusts, collaboratives and coops
  can help govern data for the maximum public benefit | nesta.
\newblock \url{https://www.nesta.org.uk/blog/new-ecosystem-trust/}.
\newblock (Accessed on 08/21/2020).

\bibitem[Narayanan(2001)]{narayanan2001water}
Vasudha Narayanan.
\newblock Water, wood, and wisdom: Ecological perspectives from the hindu
  traditions.
\newblock \emph{Daedalus}, 130\penalty0 (4):\penalty0 179--206, 2001.

\bibitem[Palinkas et~al.(2015)Palinkas, Horwitz, Green, Wisdom, Duan, and
  Hoagwood]{palinkas2015purposeful}
Lawrence~A Palinkas, Sarah~M Horwitz, Carla~A Green, Jennifer~P Wisdom, Naihua
  Duan, and Kimberly Hoagwood.
\newblock Purposeful sampling for qualitative data collection and analysis in
  mixed method implementation research.
\newblock \emph{Administration and policy in mental health and mental health
  services research}, 42\penalty0 (5):\penalty0 533--544, 2015.

\bibitem[Rangarajan et~al.(2014)Rangarajan, Mahendra~Dev, Sundaram, Vyas, and
  Datta]{Rangarajan2014Poverty}
C.~Rangarajan, S.~Mahendra~Dev, K.~Sundaram, Mahesh Vyas, and K.L Datta.
\newblock Report of the expert group to review the methodology for measurement
  of poverty.
\newblock Technical report, Government of India Planning Commission, 2014.

\bibitem[Region()]{WorldBank2009Disabilities}
World Bank Human Development Unit South~Asia Region.
\newblock People with disabilities in india: From commitments to outcomes.

\bibitem[Roberts(2016)]{roberts2016digital}
Sarah~T Roberts.
\newblock Digital refuse: Canadian garbage, commercial content moderation and
  the global circulation of social media’s waste.
\newblock \emph{Wi: journal of mobile media}, 2016.

\bibitem[RT(2017)]{Whoeverl94:online}
RT.
\newblock 'whoever leads in ai will rule the world’: Putin to russian
  children on knowledge day — rt world news.
\newblock \url{https://www.rt.com/news/401731-ai-rule-world-putin/}, September
  2017.
\newblock (Accessed on 09/20/2020).

\bibitem[Ruhaak()]{MozillaF26:online}
Anouk Ruhaak.
\newblock Mozilla foundation - when one affects many: The case for collective
  consent.
\newblock
  \url{https://foundation.mozilla.org/en/blog/when-one-affects-many-case-collective-consent/}.
\newblock (Accessed on 08/21/2020).

\bibitem[Sambasivan et~al.(2018)Sambasivan, Checkley, Batool, Ahmed, Nemer,
  Gayt{\'a}n-Lugo, Matthews, Consolvo, and Churchill]{sambasivan2018privacy}
Nithya Sambasivan, Garen Checkley, Amna Batool, Nova Ahmed, David Nemer,
  Laura~Sanely Gayt{\'a}n-Lugo, Tara Matthews, Sunny Consolvo, and Elizabeth
  Churchill.
\newblock " privacy is not for me, it's for those rich women": Performative
  privacy practices on mobile phones by women in south asia.
\newblock In \emph{Fourteenth Symposium on Usable Privacy and Security
  ($\{$SOUPS$\}$ 2018)}, pages 127--142, 2018.

\bibitem[Sambasivan et~al.(2019)Sambasivan, Batool, Ahmed, Matthews, Thomas,
  Gayt{\'a}n-Lugo, Nemer, Bursztein, Churchill, and
  Consolvo]{sambasivan2019they}
Nithya Sambasivan, Amna Batool, Nova Ahmed, Tara Matthews, Kurt Thomas,
  Laura~Sanely Gayt{\'a}n-Lugo, David Nemer, Elie Bursztein, Elizabeth
  Churchill, and Sunny Consolvo.
\newblock " they don't leave us alone anywhere we go" gender and digital abuse
  in south asia.
\newblock In \emph{proceedings of the 2019 CHI Conference on Human Factors in
  Computing Systems}, pages 1--14, 2019.

\bibitem[Thomas(2006)]{Thomas2006general}
David~R Thomas.
\newblock A general inductive approach for analyzing qualitative evaluation
  data.
\newblock \emph{American journal of evaluation}, 27\penalty0 (2):\penalty0
  237--246, 2006.

\bibitem[Toyama(2015)]{toyama2015geek}
Kentaro Toyama.
\newblock \emph{Geek heresy: Rescuing social change from the cult of
  technology}.
\newblock PublicAffairs, 2015.

\bibitem[Tunisia(2018)]{TunisiaAI}
Tunisia.
\newblock National ai strategy: Unlocking tunisia's capabilities potential.
\newblock In \emph{AI workshop.}, 2018.

\bibitem[Ullah()]{Courttol47:online}
Mazar Ullah.
\newblock Court told design flaws led to bhopal leak | environment | the
  guardian.
\newblock \url{https://www.theguardian.com/world/2000/jan/12/1}.
\newblock (Accessed on 08/21/2020).

\bibitem[website(2020)]{JayapalJ44:online}
Jayapal website.
\newblock Jayapal joins colleagues in introducing bicameral legislation to ban
  government use of facial recognition, other biometric technology -
  congresswoman pramila jayapal.
\newblock
  \url{https://jayapal.house.gov/2020/06/25/jayapal-joins-rep-pressley-and-senators-markey-and-merkley-to-introduce-legislation-to-ban-government-use-of-facial-recognition-other-biometric-technology/},
  June 2020.
\newblock (Accessed on 07/30/2020).

\bibitem[Xaxa(2011)]{xaxa2011tribes}
Virginius Xaxa.
\newblock Tribes and social exclusion.
\newblock \emph{CSSSC-UNICEF Social Inclusion Cell, An Occasional Paper},
  2:\penalty0 1--18, 2011.

\end{thebibliography}

\newpage
\appendix

\section{Methodology details}
\label{app_methodology}

We conducted qualitative interviews with 36 expert researchers, activists, and lawyers working closely with marginalised Indian communities at the grassroots. Expert interviews are a qualitative research technique used in exploratory phases, providing practical insider knowledge, surrogacy for a broader community, and importantly \cite{bogner2009interviewing}. Our respondents were chosen from a wide range of areas to create a holistic analysis of algorithmic power in India. Respondents came from Computer Science (11), Activism (9), Law and Public Policy (6), Science and Technology Studies (5), Development Economics (2), Sociology (2), and Journalism (1). All respondents had 10-30 years of experience working with marginalised communities or on social justice. Specific expertise areas included caste, gender, labour, disability, surveillance, privacy, health, constitutional rights, and financial inclusion. 24 respondents were based in India, 2 in Europe, 1 in Southeast Asia, the rest in the USA; 25 of them self-identified as male, 10 as female, and 1 as non-binary.
In conjunction with qualitative interviews, we conducted an analysis of various algorithmic deployments and emerging policies in India, starting from 2009.
We identified and analysed various Indian news publications (\eg{}TheWire.in, Times of India), policy documents (\eg{}NITI Aayog, Srikrishna Bill), grassroots fora (\eg{}Dalit Camera, Dalitality), and prior research on automation in India. 

We recruited respondents via a combination of reaching out directly and personal contacts, using purposeful sampling \cite{palinkas2015purposeful}---\ie identifying and selecting experts with relevant experience---iterative until saturation. We conducted all interviews in English (preferred language of participants). Respondents were compensated for the study (giftcards of 100 USD, 85 EUR, and 2000 INR). Interviews lasted an hour each, and were conducted using video conferencing. 
Transcripts were coded and analyzed for patterns using an inductive approach~\cite{Thomas2006general}. From a careful reading of the transcripts, we developed categories and clustered excerpts, conveying key themes from the data. Two team members created a code book based on the themes. The three power structures that we describe below were then developed and applied iteratively to the codes.

We took great care to create a research ethics protocol to protect respondent privacy and safety, especially due to the sensitive nature of our inquiry. During recruitment, participants were informed of the purpose of the study, the question categories, and researcher affiliations. Participants signed informed consent acknowledging their awareness of the study purpose and researcher affiliation prior to the interview. At the beginning of each interview, the moderator additionally obtained verbal consent. We stored all data in a private Google Drive folder, with access limited to our team. To protect participant identity, we deleted all personally identifiable information in research files. We redact identifiable details when quoting participants. Every respondent was given the choice of default anonymity or being included in acknowledgements. 
%
All co-authors of this paper work at the intersection of under-served communities and technology, with backgrounds in HCI, critical data studies, and algorithmic fairness; some of us have had grassroots commitments with marginalised Indian communities for over a decade. The first author constructed the research approach, and both the first and second author moderated interviews. All researchers were involved in the research framing, data analysis, and synthesis. Three of us are Indian and two of us are Caucasian. All of us come from privileged positions of class and/or caste. 

\newpage
\section{Axes of Discrimination in India}
\label{app_subgroups_table}

In Table~\ref{app_subgroups_table}, we describe some of the prominent axes of injustice in India that were brought up in our interviews, enriched through secondary research and statistics from authoritative sources, to substantiate attributes and proxies to substantiate attributes and proxies.
\begin{table}[h]
\small
\setlength{\tabcolsep}{2pt}
\begin{tabularx}{\linewidth}{X}
\toprule
\multicolumn{1}{c}{\bf Sub-groups, Proxies and Harms} \\
\midrule
\textbf{Caste} 
\footnotesize{(17\% Dalits; 8\% Adivasi; 40\% Other Backward Class (OBC))\cite{CensusIndia2011}} 
\begin{itemize}[noitemsep,leftmargin=*,topsep=2pt]
   \item \footnotesize{Societal harms: Human rights atrocities. Poverty. Land, knowledge and language battles  \cite{xaxa2011tribes, ambedkar2014annihilation, Twothird31:online}. }
    \item \footnotesize{Proxies: Surname. Skin tone. Occupation. Neighborhood. Language.} 
    \item \footnotesize{Tech harms: Low literacy and phone ownership. Online misrepresentation \& exclusion. Accuracy gap of Facial Recognition {(FR)}. Limits of Fitzpatrick scale. Caste-based discrimination in tech ({}\cite{TheCisco63:online}). }
\end{itemize}
\\ \midrule
\textbf{Gender}
\footnotesize{(48.5\% female)\cite{IndiaMinistryStatistics2018}} 
\begin{itemize}[noitemsep,leftmargin=*,topsep=2pt]
    \item \footnotesize{Societal harms: Sexual crimes. Dowry. Violence. Female infanticide.}
  \item \footnotesize{Proxies: Name. Type of labor. Mobility from home. }%
    \item \footnotesize{Tech harms: Accuracy gap in FR. Lower creditworthiness score. Recommendation algorithms favoring majority male users. Online abuse and 'racey’ content issues. Low Internet access.} 
\end{itemize}
\\ \midrule

\textbf{Religion}
\footnotesize{(80\% Hindu, 14\% Muslim, 6\% Christians, Sikhs, Buddhists, Jains and indigeneous) \cite{CensusIndia2011}}  
\begin{itemize}[noitemsep,leftmargin=*,topsep=2pt]
    \item \footnotesize{Societal harms: Discrimination, lynching, vigilantism, and gang-rape against Muslims and others \cite{84DeadIn98:online}.}
    \item \footnotesize{Proxies: Name. Neighborhood. Expenses. Work. Language. Clothing. } 
    \item \footnotesize{Tech harms: Online stereotypes and hate speech, \eg Islamophobia. Discriminatory inferences due to lifestyle, location, appearance. Targeted Internet disruptions.} 
\end{itemize}
\\ \midrule

\textbf{Ability}
\footnotesize{(5\%--8\%+ persons with disabilities) \cite{WorldBank2009Disabilities}} 
\begin{itemize}[noitemsep,leftmargin=*,topsep=2pt]
    \item \footnotesize{Societal harms: Stigma. Inaccessible education, transport \& work.} 
    \item \footnotesize{Proxies: Non-normative facial features, speech patterns, body shape \& movements. Use of assistive devices.} 
    \item \footnotesize{Tech harms: Assumed homogeneity in physical, mental presentation. Paternalistic words and images. No accessibility mandate. }
\end{itemize}
\\ \midrule

\textbf{Class}
\footnotesize{(30\% live below poverty line; 48\% on \$2--\$10/day)\cite{Rangarajan2014Poverty} \cite{CensusIndia2011}}
\begin{itemize}[noitemsep,leftmargin=*,topsep=2pt]
    \item \footnotesize{Societal harms: Poverty. Inadequate food, shelter, health, \& housing.}
    \item \footnotesize{Proxies: Spoken \& written language(s). Mother tongue. Literacy. Feature / Smart Phone Ownership. Rural vs. urban.}
    \item \footnotesize{Tech harms: Linguistic bias towards mainstream languages. Model bias towards middle class users. Limited or lack of internet access. }
    \end{itemize}
\\ \midrule

\textbf{Gender Identity \& Sexual Orientation}
\footnotesize{(No official LGBTQ+ data)} 
\begin{itemize}[noitemsep,leftmargin=*,topsep=2pt]
    \item \footnotesize{Societal harms: Colonial law 377 traces. Discrimination and abuse.} 
    \item \footnotesize{Proxies: Gender declaration. Name. }%
    \item \footnotesize{Tech harms: FR "outing" and accuracy. Gender binary surveillance systems (\eg in dormitories). M/F ads targeting.  Catfishing and extortion abuse attacks.}
\end{itemize}
\\ \midrule

\textbf{Ethnicity}
\footnotesize{(4\% NorthEast) \cite{CensusIndia2011}}
\begin{itemize}[noitemsep,leftmargin=*,topsep=2pt]
    \item \footnotesize{Societal harms: Racist slurs, discrimination, and physical attacks.}
    \item \footnotesize{Proxies: Skin tone. Facial features. Mother tongue. State. Name. }
    \item \footnotesize{Tech harms: Accuracy gap in FR. Online misrepresentation \& exclusion. Inaccurate inferences due to lifestyle, \eg migrant labor. }
\end{itemize}
\\ \bottomrule
\end{tabularx}
\normalsize
\caption{\label{app_subgroups_table} \small Axes of potential ML (un)fairness in India: sub-groups, proxies, and harms.
}
\end{table}

\end{document}